\documentclass[twocolumn]{autart}
 
\input epsf

\usepackage{graphicx,amsmath,amssymb,remark,color}
 \usepackage{algorithm} 
\usepackage{algorithmic}

\usepackage{epsfig}
\newcommand{\EQ}{\begin{eqnarray}}
\newcommand{\EN}{\end{eqnarray}}
\newcommand{\EQQ}{\begin{eqnarray*}}
\newcommand{\ENN}{\end{eqnarray*}}
\newcommand{\R}{\mathbb R}

\usepackage[outdir=./]{epstopdf}
\graphicspath{{../eps/}}
\DeclareGraphicsExtensions{.eps}

\newcommand{\bremark}
{\medskip\begin{remark}
\begin{rm}}
\newcommand{\eremark}{ \end{rm}\hfill \rule{1mm}{2mm}
\end{remark} }
\newcommand{\btheorem}{\medskip\begin{theorem} \begin{it}}
\newcommand{\etheorem}{\end{it} \hfill \rule{1mm}{2mm}
\end{theorem} }
\newcommand{\blemma}{\medskip\begin{lemma} \begin{it} }
\newcommand{\elemma}{ \end{it} \hfill\rule{1mm}{2mm}
\end{lemma} }
\newcommand{\bcorollary}{\medskip\begin{corollary} \begin{it} }
\newcommand{\ecorollary}{ \end{it} \hfill\rule{1mm}{2mm}
\end{corollary} }
\newcommand{\bdefinition}{\medskip\begin{definition} }
\newcommand{\edefinition}{ \hfill\rule{1mm}{2mm}
\end{definition} }
\newcommand{\bproposition}{\medskip\begin{proposition} }
\newcommand{\eproposition}{\hfill \rule{1mm}{2mm}
\end{proposition} }
\newcommand{\bexample}{\medskip\begin{example} \begin{rm}}
\newcommand{\eexample}{ \end{rm} \hfill\rule{1mm}{2mm}
\end{example} }

\newcommand{\basm}{\medskip\begin{assumption} \begin{rm} }
\newcommand{\easm}{ \end{rm} \hfill\rule{1mm}{2mm}  
\end{assumption} }
\newcommand{\bru}{\medskip\begin{rule1} \begin{rm}}
\newcommand{\eru}{ \end{rm} \hfill\rule{1mm}{2mm}
\end{rule1} }

\newtheorem{theorem}{\sf\bfseries Theorem}[section]
\newtheorem{lemma}{\sf\bfseries Lemma}[section]

\newtheorem{definition}{\sf\bfseries Definition}[section]
\newtheorem{remark}{\sf\bfseries Remark}[section]
\newtheorem{corollary}{\sf\bfseries Corollary}[section]
\newtheorem{proposition}{\sf\bfseries Proposition}[section]
\newtheorem{example}{\sf\bfseries Example}[section]
\newtheorem{assumption}{\sf\bfseries Assumption}
\newtheorem{rule1}{\sf\bfseries Rule}[section]
\newenvironment{Proof}{\noindent{\em Proof:\/}}{\hfill $\Box$\par}
\begin{document}

\begin{frontmatter}

\title{Discrete-Time Consensus Networks: Scalability, Grounding and Countermeasures
}

\thanks[footnoteinfo]{This work was supported by the Australian Research Council through grant DP190102859.}
%\thanks[footnoteinfo]{The preliminary version of  is published in the 2020IFAC World Congress.}
\author{Yamin Yan, Sonja St{\"u}dli, Maria M. Seron and Richard H. Middleton}

\address{School of Electrical Engineering and Computing\\
The University of Newcastle\\
Callaghan, NSW 2308, Australia\\
Email:  {\tt yamin.yan, sonja.stuedli, maria.seron, richard.middleton@newcastle.edu.au}}  
 
  \date{}
  
\begin{keyword}  Grounding, discrete-time systems, scalability, consensusability, consensus performance, countermeasures
\end{keyword}

 \begin{abstract}
We investigate the disruption of discrete-time consensus problems via grounding. Loosely speaking, grounding a network occurs if the state of one agent no longer responds to inputs from other agents and/or changes its dynamics. Then, the agent becomes a leader or a so-called stubborn agent. The disruption of the agent can be caused by internal faults, safety protocols or due to an external malicious attack. 
In this paper we investigate how grounding affects expander graph families that usually exhibit good scaling properties with increasing network size. It is shown that the algebraic connectivity and eigenratio of the network decrease due to the grounding causing the performance and scalability of the network to deteriorate, even to the point of losing consensusability. We then present possible countermeasures to such disruptions and discuss their practicality and limitations. In particular, for a specific countermeasure of deliberately grounding additional nodes, we investigate extensively how to select additional nodes to ground and how many nodes we need to ground to recover the consensus performance. Our findings are supported by a wide range of numerical simulations. 
\end{abstract}

\end{frontmatter}
 %%%%%%%%%%%%%%%%%%%%%%%%%%%%%%%%%%%%%%%%%%%%%
 \section{Introduction}

The multi-agent consensus problem has been a prevalent research area over the past few decades \cite{jadbabaie2003,olfati2007consensus,knorn2016}. Among the many studies on consensus, fundamental questions such as whether the network can achieve consensus (consensusability) \cite{you2011}, how to achieve consensus and consensus on what \cite{li2010consensus}, are major topics of interest. In addition, analysis of consensus performance such as convergence rate is of both  theoretical and practical importance \cite{olfati2007consensus,chen2015convergence}.   

Efficient distributed networked control requires desirable scalability and consensus performance. Scalability means preservation of stability of
the entire network as the network size grows large (addition of agents). Consensus networks with bounded nodal degree, which is desired for efficiency of communication, usually scale poorly, demonstrating scaling fragility. One type of graph family, called expander family (or expanders), scales well while keeping a bounded nodal degree \cite{pinsker1973expander}. These graphs play an important role in designing efficient communication networks. It is known that the algebraic connectivity, the second smallest eigenvalue of the graph Laplacian, is crucial in characterizing scalability and consensus performance \cite{1707883}. The algebraic connectivity of expander families with non-increasing nodal degree is bounded away from zero, thus possessing desirable scalability and consensus properties. Early studies on consensus of expanders can be found in \cite{li2009quantized}.

Recently, \cite{emma} revealed the scale fragility of expander families towards grounding in a \emph{continuous-time} setting. For discrete-time multi-agent systems, \cite{yan2020analysis} revealed not only scale fragility but also the deterioration of consensus performance and, in the worst case, even loss of consensusability. Moreover, \cite{yan2020analysis} suggested possible countermeasures   for disruption via grounding over expander networks.
Grounding means that the grounded node is no longer affected by other agents while still influencing its neighbors and, by doing so, the complete network. Another way of interpreting this behavior is that the grounded node acts as a leader, hence turning the whole network from a leaderless architecture to a leader-following one. The terminology stems from its application in power networks where grounding a node means literally connecting the bus to ground, thus forcing the state to be set to zero. Grounding can be caused by internal faults, safety protocols or due to an external attack. The latter can be viewed as disruption/deception attacks by either disconnecting the input channel or changing the dynamics \cite{dibaji2019systems}.

Once a network has been grounded, its dynamics can be described by a grounded Laplacian, also known as Dirichlet Laplacian \cite{barooah2006graph}. The study of the grounded Laplacian has recently received increasing attention. For example, \cite{pirani2015smallest,pirani2017robustness} extensively study the spectral properties of the grounded Laplacian for undirected graphs, while \cite{xia2017analysis} considers directed grounded networks. In discrete-time settings, the factor called eigenratio \cite{you2011}, i.e., the ratio of the second smallest to the largest eigenvalue of the Laplacian,  plays a significant role in characterizing consensusability of undirected graphs. It was shown in \cite{yan2020analysis} that, for expander graph networks, while the eigenratio of the nongrounded graph is bounded away from zero with increasing network size, this no longer holds for the grounded graph. For unstable system dynamics, this reduction of the eigenratio impacts on consensusability, which in the worst case can be lost.

In this paper, we adopt a scalable design for the consensus protocol as well as the communication graph structure from \cite{sonja2020ANZCC}. The controller design makes use of a lower bound of the algebraic connectivity and is independent of the number of agents. For this design we then investigate the impact of grounding a node. We find that instability in the case of unstable agent dynamics is unavoidable for the grounded system as the number of agents grows large enough if no countermeasure is taken. Finally, we present possible countermeasures and give insights on their practicality and their limitations.

The contributions of this paper are three-fold. Firstly, we investigate the properties of scalability, consensus performance and consensusability of non-regular expander networks towards grounding, which extends the result in \cite{yan2020analysis} applied to regular networks only. Secondly, we expand countermeasures in \cite{yan2020analysis} to mitigate the undesirable fragility over grounding in both passive and active manners. Thirdly, we extensively study the specific countermeasure of grounding additional nodes on how to effectively select the additional nodes and how many nodes need to be grounded to recover the performance. Two practical algorithms are proposed for the selection of the additional nodes. 

 The remainder of this paper is organized as follows. In Section~2, we provide network graph notation and definitions, as well as background information on discrete-time learderless (nongrounded) consensus networks. In Section~3, the leader-following (grounded) consensus problem with a summary of network properties that this paper analyzes are presented. In Section~4, we discuss the scaling fragility and loss of consensusability towards grounding. In Section~5, possible countermeasures to mitigate the detrimental effects of grounding are summarized. We study the countermeasure of grounding additional nodes in Section~6. To illustrate our analysis on grounding, three numerical simulations are given in Section~7. Finally, the paper is closed with some concluding remarks in Section~8. 
%We then show the deterioration and even loss of consensusability caused by grounding in Section~4 and 
\section{Preliminaries}
\subsection{Network Graph Definitions}
A graph is denoted by $\mathcal{{G}}=(\mathcal{{V}},\mathcal{{E}})$, where $\mathcal{{V}}=\{1,\cdots,N\}$ is the node set and $\mathcal{{E}}$ is the edge set. A graph is undirected if the edge set consists of unordered pairs $(i,j) \in \mathcal{E}, i,j=1,\cdots,N$, if there is communication between nodes $i$ and $j$. A graph is called simple if there are no loops ($(i,i) \not\in \mathcal{E}~ \forall i \in \mathcal{V}$) and each edge is present only once in $\mathcal{E}$. A graph family $\{\mathcal{G}_N\}$ is a sequence of graphs with increasing number of nodes, i.e., $N \rightarrow \infty$.

If $(i,j) \in \mathcal{{E}}$, node $j$ is called a neighbor of node $i$. $\mathcal{N}_i$ denotes the neighbor set of node $i$ with respect to $\mathcal{V}$. The degree of a node $i$ is equal to the number of neighbors node $i$ has, i.e., $d_i=|\mathcal{N}_i|$. The degree matrix $\mathcal{D}$ is a diagonal matrix having the elements $d_i$ on the diagonal. A graph is called $d$-regular if the degree of each node is equal to $d$, and is called non-regular if it is not a regular graph.

Let $\mathcal A=[\alpha_{ij}]_{i,j=1}^N\in\R^{N\times N}$ be the weighted adjacency matrix of ${\mathcal{G}}$ with $i,j = 1, \cdots, N$,  $\alpha_{ii}=0$, and  $\alpha_{ij}=\frac{1}{d_i}$ $\Leftrightarrow$ $(j,i)\in \mathcal{ E}$. The random walk normalised Laplacian matrix $L=[l_{ij}]\in\R^{N\times N}$ associated with $\mathcal{G}$ is defined as $l_{ii}=1$ and $l_{ij}=-\alpha_{ij}, i\neq j$, we have $L=I_N-\mathcal{A}$. The symmetric normalized adjacency matrix is defined as $\mathcal{A}_{sym}=\mathcal{D}^{\frac{1}{2}}\mathcal{A}\mathcal{D}^{-\frac{1}{2}}$ and its associated Laplacian matrix $L_{sym}$ is denoted by $L_{sym}=\mathcal{D}^{\frac{1}{2}}L\mathcal{D}^{-\frac{1}{2}}=I_N-\mathcal{A}_{sym}$. 

%  An important property for graphs is how well they are connected. There are several measures that capture some aspect of connectedness. In this paper we use the second smallest eigenvalue of the normalised Laplacian as a measure. This measure is often called the algebraic connectivity. 

%  Let $\chi_{1} < \chi_{2}\leq\ldots \leq \chi_{N}$ be the eigenvalues of the normalised  Laplacian $L$. It is well known that $\chi_{1} = 0$ and  $\chi_{i}<2$ for $i \in \set{1,2,\ldots,N}$. Further, $\chi_{2}>0$ if and only if the graph is connected. We define the following term for a graph.
Throughout this paper, we consider undirected, simple, and connected communication graphs. The eigenvalues of $L_{sym}$ and $L$ agree since they are similar and are denoted by $\lambda_{i} \in \R,\ i=1,\cdots,N$ and in an ascending order are written as $0 = \lambda_{1} \leq \lambda_{2} \leq \cdots \leq \lambda_{N}$ and $\lambda_i\leq2$ for $i=2,\dots,N$. We denote $\lambda_{\max}(\boldsymbol{M})$ and $\lambda_{\min}(\boldsymbol{M})$ by the largest and the smallest positive eigenvalue of a symmetrical matrix $\boldsymbol{M}$, respectively; see e.g., $\lambda_{\min}(L)=\lambda_2$. The second smallest eigenvalue of the Laplacian matrix, $\lambda_2$, is called the algebraic connectivity of the graph. As in \cite{you2011}, $\rho=\lambda_{2}/\lambda_{N}$, is called the eigenratio of an undirected graph. We consider throughout the paper graph families, $c^{\prime}$-expanders, with the so-called minimum $c^{\prime}$-algebraic connected property.
%This means that the weighted adjacency matrix and the normalised Laplacian matrix $L$ are symmetric.
\begin{definition}\cite{sonja2020ANZCC}
  \label{thm:algebraic-expander-def}
  Let $c'$ be a positive constant. A graph is called minimum $c'$-algebraic connected if $\lambda_{2} \geq c'$.
\end{definition}

\begin{definition}\label{c'connected}
A graph family $\{{\mathcal{G}}_N\}$ is called $c^{\prime}$-expanders if every $\mathcal{G}$ in $\{{\mathcal{G}}_N\}$ satisfies the minimum $c'$-algebraic connected condition.
\end{definition}
 Generally, as the number of nodes increases while keeping a constant nodal degree, the algebraic connectivity tends towards zero. This means that with increasing $N$ the graph loses its connectivity and the performance of the consensus algorithm deteriorates. A minimum $c^{\prime}$-algebraic connected graph family does not exhibit this decrease in connectivity while maintaining a bounded nodal degree.
% \bdefinition
%\cite{krebs2011expander} (Expander family) Let $\{\mathcal{G}_N\}$ be a graph family in which $N\rightarrow \infty$. If the sequence $h(\mathcal{G}_N)$ is bounded away from $0$, $\{\mathcal{G}_N\}$ is an expander family.
%\edefinition

% Note that since in an expander family $h(\mathcal{G}_{N})$ is bounded away from $0$, so is the algebraic connectivity $\lambda_{2}$ and the eigenratio $\frac{\lambda_{2}}{\lambda_{N}}$ due to the Cheeger inequality, see \eqref{eq:cheeger-inequality} and \eqref{eq:eigenratio-bound}. 

\subsection{Multi-agent consensus network}
We consider a discrete-time multi-agent system where $N$ agents communicate among each other to achieve consensus on their states in a leader-less architecture. Each agent is governed by a discrete-time dynamic system given in the form of
\EQ\label{sys1}
x_i(k+1)=Ax_i(k)+Bu_i(k), k\in \mathbb{Z}^+, ~i=1,\dots,N
\EN
where $x_i\in\R^n$, $u_i\in \R$ denote the system state and control input of the $i^{th}$ agent, $A\in \R^{n\times n}$, $B\in \R^{n\times 1}$. $\mathbb{Z}^+$ denotes the set of nonnegative integers $\mathbb{Z}^+ =  \{0, 1, \cdots \}$. The standard consensus algorithm is adopted as follows,
\EQ\label{con}
u_i(k)=K\sum_{j\in\mathcal{N}_i}\alpha_{ij}(x_j(k)-x_i(k))
\EN
where $K\in \R^{1\times n}$ is the control gain matrix, $\alpha_{ij}$ the $ij$-th entry of the adjacency matrix $\mathcal{A}$.

 The multi-agent consensus network composed of \eqref{sys1} and \eqref{con} can be expressed in the following compact form
 \EQ\label{cl}
x(k+1)=(I_N\otimes A-L\otimes BK)x(k)
 \EN
where $x=[x_1^T,\dots,x_N^T]^T$ and $\otimes$ denotes the Kronecker product.

Throughout the paper we assume the following. 

\basm\label{asm2}
The pair $(A, B)$ is controllable and 
\EQ \label{Auinequal}
 \tilde \sigma:=\frac{1}{\prod_j|\lambda_j^u(A)|}>\frac{1-\rho}{1+\rho}
\EN
where $\lambda_j^u(A)$ is an unstable eigenvalue of $A$, the product in \eqref{Auinequal} is over all such eigenvalues, and $\tilde \sigma = 1$ if $A$ is stable.
\easm

It is known that a necessary and sufficient condition for consensus of system~\eqref{cl} is that there exists $K$ such that $A-\lambda_iBK$ is Schur (i.e. all its eigenvalues are inside the open unit circle) for $i=2,\dots,N$ \cite{you2011}.
 Under Assumption~\ref{asm2}, consider a minimum $c^{\prime}$-algebraic connected graph family where $c^{\prime}$ should satisfy as in \cite{sonja2020ANZCC} $$\frac{2(1-\tilde \sigma)}{1+\tilde \sigma}<c^{\prime}<2.$$  Then, we design 
 \EQ\label{K}
 K=\epsilon\frac{B^TPA}{B^TPB+R}
 \EN where $R\geq 0$, $P=P^T>0$ is a solution to the modified algebraic Riccati inequality
\EQ\label{MARI}
P-A^TPA+(1-\sigma^2)\frac{A^TPBB^TPA}{B^TPB+R}>0,
\EN 
with $\sigma$ satisfying 
\EQ \label{sigma}
\frac{2-c'}{2+c'}\leq\sigma<\tilde{\sigma}.
\EN
From \cite{sonja2020ANZCC}, we select $\epsilon$ to satisfy
\EQ\label{epsilon}
\frac{1-\sigma}{c^{\prime}}\leq\epsilon\leq\frac{1+\sigma}{2},
\EN 
which guarantees that the gain $K$ in \eqref{K} is such that the network achieves consensus.

\bremark
It is worth pointing out that \eqref{Auinequal} in Assumption~1 is a necessary and sufficient condition for consensusability using the design in \cite{you2011} with $K=\frac{2}{\lambda_2+\lambda_N}\frac{B^TPA}{B^TPB}$, where $P$ is the solution to \eqref{MARI} with $\sigma$ satisfying $\frac{1-\rho}{1+\rho}<\sigma\leq \tilde \sigma$. We can see that the left-hand side $\frac{2-c'}{2+c'}$ in \eqref{sigma} is greater than or equal to the right-hand side $\frac{1-\rho}{1+\rho}$ in \eqref{Auinequal}. Only when the algebraic connectivity $\lambda_2$ equals $c'$ and the spectral radius $\lambda_N$ equals $2$ can we have $\frac{1-\rho}{1+\rho}=\frac{2-c'}{2+c'}$. This also means that $\frac{2-c'}{2+c'}$ is a stricter bound for $\tilde \sigma$. However, the design in \cite{you2011} uses explicitly $\lambda_2$ and $\lambda_N$ of the communication graph while the design in \eqref{K} does not require reconfiguration when the network size changes for a $c^{\prime}$-expander graph family.
\eremark
\section{Grounding and Problem Formulation}

In this section, we first introduce the concept of grounding and the grounded (leader-following) consensus network as in \cite{yan2020analysis} for the consideration of being self-contained, then we summarize the properties and results that this paper analyzes and establishes for these problems. %We then summarize the properties and results that this paper analyzes and establishes for these problems. 

\subsection{Grounded networks (leader-following architecture)}
% {\blue Sonja:  }

Grounding a node of a network turns the node into one that influences other nodes but is not affected in return. In a multi-agent context, this grounded node acts as a leader and converts the whole network from a leaderless architecture to a leader-following one. In other contexts, the grounded node can be interpreted as a ``stubborn agent''\cite{ghaderi2013opinion}. As mentioned previously, the terminology stems from its application in power networks where grounding a node means literally connecting the bus to ground forcing the state, in this case the voltage, to be set to $0$. To put this concept in a general framework using networked control language, we consider three different ways to ground a node. These different ways will have different influences on network consensus. Without loss of generality, we suppose grounding happens to node 1.

%in the Introduction, the concept of grounding comes from the literature in power networks where a node is connected to ground so the voltage level becomes zero for reference.

A first form of grounding consists in fixing node 1's state $x_1(k)$ at some time $k_0$ to be either its current state or any constant value $\bar c$ in the proper dimension. Then, $x_1(k)=\bar c$ for all $k\geq k_0$. The closed-loop
system for the remaining nodes can be described as
\EQ\label{sysgrounded1}
\bar x(k+1)=&&(I_{N-1}\otimes A-\bar{L}\otimes BK)\bar x(k)\nonumber\\&&+(\Lambda\otimes BK)(\mathbf{1}_{N-1}\otimes \bar c)
\EN
where $\bar L$ is the grounded Laplacian \cite{barooah2006graph} obtained by removing the
first row and column of $L$, $\Lambda$ is a diagonal matrix with diagonal entries equal to $\alpha_{i1}$ and $\bar x$ is obtained from $x$ by removing
the states of node 1. If $(I_{N-1}\otimes A-\bar{L}\otimes BK)$ is Schur, then $\bar x$ will approach $(I_{n(N-1)}-(I_{N-1}\otimes A-\bar{L}\otimes BK))^{-1}(\Lambda\otimes BK)(1_{N-1}\otimes \bar c)$. 

A second form of grounding consists in cutting the control channel so that $u_1=0$, and optionally change the dynamics of node 1. Then, the first node's dynamics are $x_1(k+1)=\bar A x_1(k)$. If $\bar A=A$, the consensus trajectory will be the same as that of $x_1$ if grounding happens after the consensus was achieved; the consensus trajectory will be different from that of $x_1$ if grounding happens before the consensus is achieved. %{\blue can mean to cut of the control channel and optionally change the dynamics of agent 1. Then, the first agent's dynamics are $x_1(k+1) = \bar{A} x_1(k)$.}

A third form of grounding consists in taking control of $u_1$ such that $x_1$ is steered towards a deliberately designed trajectory, for example, a certain setpoint $c_0$. Specifically, a stabilizing controller $u_1=-K_1x_1+c_1$ makes the closed-loop dynamics of node 1 $x_1(k+1)=(A-BK_1)x_1(k)+Bc_1$ with $(A-BK_1)$ Schur. The closed-loop
system for the remaining nodes will be 
\EQ\label{sysgrounded3}
\bar x(k+1)=&&(I_{N-1}\otimes A-\bar{L}\otimes BK)\bar x(k)\nonumber\\&&+(\Lambda\otimes BK)(\mathbf{1}_{N-1}\otimes x_1(k)).
\EN
If $(I_{N-1}\otimes A-\bar{L}\otimes BK)$ is Schur, all states of the remaining nodes will approach $ c_0=(I-(A-BK_1))^{-1}Bc_1$.  

Note that to analyze the consequences of grounding, a key system matrix to be analyzed is $(I_{N-1}\otimes A-\bar{L}\otimes BK)$. Actually, when letting $x_1=\mathbf{0}_n$, where $\mathbf{0}_n$ denotes the column vector of $n$ zeros, the closed-loop
system for the remaining nodes will be
\EQ\label{sysgrounded}
\bar x(k+1)=(I_{N-1}\otimes A-\bar{L}\otimes BK)\bar x(k).
\EN
The performance of the grounded network will be directly related to the grounded Laplacian $\bar L$. We denote the eigenvalues of $\bar L$ as $\bar \lambda_i$ and they are numbered as $0<\bar\lambda_1\leq\dots\leq\bar \lambda_{N-1}$.
The smallest eigenvalue $\bar \lambda_1$ is known as grounded algebraic connectivity. We denote the ratio $\bar\rho=\bar \lambda_1/\bar \lambda_{N-1}$ by grounded eigenratio.
\subsection{Problem formulation}
In the following sections, we present results and discussions on spectral properties of $\bar L$ in regard to scalability, consensus performance, and consensusability.  Specifically, we address the following problems. 
%
%{\blue 1) scalability of consensus: consensus is always achieved no matter what size the network is
%
%2) scalability of consensus performance: convergence rate to consensus is (somewhat) independent of network size}
%

\emph{Scalability, performance, and consensusability:} Suppose the network graph has good scaling properties, that is, it is possible to achieve consensus as the network size grows large. Is this property preserved when a node is grounded. Is the convergence rate to consensus afffected by grounding. And, also how is the consensusability condition of Assumption 1 affected when the system dynamics are unstable. Can consensusability be lost.

\emph{Possible countermeasuers:} When interpreting grounding as an attack, what are the possible countermeasures that can be taken to correct or minimise the effect of grounding. How to choose an optimal countermeasure to recover the performance or in the worst case the consensusability. One countermeasure we will in particular investigate is the grounding of additional nodes. We investigate how to select effectively which nodes to ground and how many nodes should be grounded to recover the consensus performance.

%\emph{Scalability of consensus:} Suppose the network graph has good scaling properties, that is, it is possible to achieve consensus as the network size grows large (addition of agents). Is scalability preserved when a node is grounded. 
%
%
%\emph{Consensus performance:} Is the convergence rate to consensus affected by grounding.
%
%
%\emph{Consensusability for unstable systems:} How is the consensusability condition of Assumption 1 affected by grounding and can consensusability be lost after grounding.
%
%In addition, when interpreting grounding as an attack, what are the possible countermeasures that can be taken to correct or minimise the effect of grounding. How to choose an optimal countermeasure to recover the consensus performance or the consensusability. 
%
%%\emph{
%In particular, as a countermeasure by deliberately grounding additional nodes, how to select wisely which node to ground and how many nodes should be grounded to recover the consensus performance. 
%%
%
\section{Scalability and Consensusability of Grounded Networks}
The advantages of using expander graphs for the consensus algorithm are clear from the previous section. These advantages are not perserved when grounding happens to the network. 

First and foremost the scalability of the grounded network is limited. While $\lambda_2$ is bounded away from $0$ due to the property of the expander family, the larger lower bound of $\lambda_2$, the better the system scaling tends to be. However, in the grounded network, $\bar{\lambda}_1$ approaches zero as $N$ grows. This means that the scalability is limited.

Secondly the performace (convergence rate) of the grounded network degrades. The convergence rate directly depends on the algebraic connectivity \cite{olfati2004consensus,olfati2007consensus}. A lower algebraic connectivity indicates a slower convergence of the consensus algorithm. 

The following result presents the scaling fragility and performance degradation by showing the upper bound of $\bar {\lambda}_1$ approaching $0$ as the network size grows and hence $\lambda_2 > \bar{\lambda}_1$ for large enough $N$. It is worth mentioning that compared to \cite{yan2020analysis}, the following lemma provides a more general result for non-regular communication graphs.

\blemma\label{lemma_1_bound}
Consider a Laplacian matrix $L$ of an undirected network with bounded nodal degree. Let $d_{\max}$ and $d_{\min}$ be the maximum and minimum degree, respectively, among all the $N$ nodes. The smallest eigenvalue of the grounded Laplacian $\bar L$  is bounded such that
\EQ \label{bar_lambda1}
    \bar {\lambda}_{1} \leq \frac{d_{\max}}{(N-1) ~d_{\min}}.
\EN
\elemma
\begin{Proof}
%To prove Lemma~\ref{lemma_1_bound}, we note that the normalized Laplacian $L$ is similar to the symmetric normalized Laplacian $L_{sym}$ given by $L=\mathcal{D}^{-\frac{1}{2}}L_{sym}\mathcal{D}^{\frac{1}{2}}$ which means that the spectrum of $L$ and $L_{sym}$ are the same.%, thus the smallest eigenvalue of $L_{sym}$ is equal to $\bar\lambda_1$. 
We denote $\bar L_{sym}$ the grounded symmetric normalized Laplacian which is obtained by removing the first row and column of $L_{sym}$. Also, the Laplacian can be partitioned as follows
\EQ
L=\left[\begin{array}{cc}
1&\hat L_0\\
\tilde L_0& \bar L
\end{array}\right]
\EN
Using Ritz-Rayleigh theorem, one has
\EQ\label{lambda_m}
\bar{\lambda}_1 \leq \frac{y^T\bar L_{sym} y}{y^Ty}, ~y\neq 0.
\EN
We consider a vector $s=\left[\sqrt{d_{2}},~\sqrt{d_{3}},\dots,~\sqrt{d_N}\right]^T$. Note that $s^Ts=\sum_{i=2}^Nd_i$ and we find that
\EQ
(N-1)d_{\min}\leq s^Ts\leq (N-1)d_{\max}.
\EN
Then, \eqref{lambda_m} holds when $y=s$ with $\bar\lambda_1 \leq \frac{s^T\bar L_{sym} s}{s^Ts}$.

Note that the $(i-1)$-th element of $\bar L_{sym}s$ is given as
\EQ
\sqrt{d_i}-\sum_{j=2}^N\sqrt{\alpha_{ij}}
\EN
since  $\alpha_{ij}=\frac{1}{d_i}$ for $(i,j)\in\mathcal{E}$ and $0$ otherwise, $(i-1, j-1)$th entry of $\bar L_{sym}$ is $-\frac{1}{\sqrt{d_i d_j}}$ for $(i,j) \in\mathcal{E}$, and $0$ otherwise.

We next note that $\sum_{j=2}^N\sqrt{\alpha_{ij}}=\sqrt{d_i}-\sqrt{\alpha_{i1}}$, which can be obtained by remarking that $\frac{d_i}{\sqrt{d_i}}$ is the $i$th row sum of $\sqrt{\mathcal{A}}$, where the square root is taken element wise. Thus the $(i-1)$-th element of $\bar L_{sym} s$ is equal to $\sqrt{\alpha_{i1}}$.

Using the above yields $s^T\bar L_{sym}s=s^T\sqrt{-\tilde L_0}$, with the square root taken element wise.% and $\mathbf{1}$ is the vector of all ones with $m$ entries.

Next we find 
$s^T\bar L_{sym}s =  d_1$
by remarking that $s^T\sqrt{-\tilde L_0} = d_1$, since the $(i-1)$-th element of $\tilde L_0$ is equal to $-\alpha_{1j}$, and $\tilde L_0$ has $d_1$ nonzero elements.

Hence, we find 
\EQQ
\bar \lambda_1\leq \frac{s^T\bar L_{sym}s}{s^Ts}\leq \frac{d_1}{(N-1)d_{\min}}\leq \frac{d_{\max}}{(N-1)d_{\min}}.
\ENN
\end{Proof}
\bremark
It is noted from above that the upper bound $\frac{d_{1}}{(N-1)d_{\min}}$ is stricter than  $\frac{d_{\max}}{(N-1)d_{\min}}$, where $d_1$  denotes the degree of the grounded node. It presents a relaxed bound in \eqref{bar_lambda1} as the degree of the grounded node may not always be known.  
\eremark
By Lemma~\ref{lemma_1_bound}, the following result follows.
\blemma \label{lemma1}
Consider a Laplacian matrix $L$ and its grounded Laplacian matrix $\bar L$, of an undirected minimum $c^{\prime}$-algebraic connected graph family $\{\mathcal{G}_N\}$, then, there exists a network size $\bar N=1+\frac{d_{\max}}{c^{\prime}d_{\min}}$ such that $\lambda_2>\bar \lambda_1$, for $N>\bar N$. % $$\sigma(L)=\{0,\lambda_2,\dots,\lambda_N\}$$ where $0<\lambda_2\leq\dots\leq\lambda_N$. $\bar L$ is the grounded Laplacian obtained by removing the first row and column (WLOG) of $L$ with $$\sigma(\bar L)=\{\bar \lambda_1,\dots,\bar\lambda_{N-1}\},$$ where $0<\bar\lambda_1\leq\dots\leq\bar\lambda_{N-1}$. Then, $\lambda_2>\bar \lambda_1$.
\elemma
\begin{Proof}
By Lemma~\ref{lemma_1_bound}, $\bar \lambda_1\to 0$ as $N\to\infty$. %we  gives the bound of $\bar \lambda_1$, we then know that
Since $\lambda_2\geq c^{\prime}$, there exists $\bar N=1+\frac{d_{{\max}}}{c^{\prime}d_{\min}}$, such that  for $N>\bar N$
\EQ
\bar \lambda_1\leq \frac{d_{{\max}}}{(N-1)d_{\min}}<\frac{d_{{\max}}}{(\bar N-1)d_{\min}}=c',
\EN
hence $\lambda_2>\bar \lambda_1$.
\end{Proof}

It is known that the eigenratio $\frac{\lambda_2}{\lambda_N}$ is an important factor in discrete-time networks. A larger eigenratio corresponds to better consensusability of the communication graph. Grounding causes degradation of consensus and can even destroy consensusability. We investigate in the following when this occurs. Our argument is based on the observation that the eigenratio of the grounded network is smaller than that of the nongrounded network for large $N$. We then have the following result.

\blemma \label{lemma2}
% For a Laplacian matrix $L$ of an undirected connected regular graph $\mathcal{G}$, $$\sigma(L)=\{0,\lambda_2,\dots,\lambda_N\}$$ where $0<\lambda_2\leq\dots\leq\lambda_N$. $\bar L$ is the grounded Laplacian obtained by removing the first row and column (WLOG) of $L$ with $$\sigma(\bar L)=\{\bar \lambda_1,\dots,\bar\lambda_{N-1}\},$$ where $0<\bar\lambda_1\leq\dots\leq\bar\lambda_{N-1}$. Then,\\

% 1) $\lambda_2>\bar \lambda_1$; \\

Consider a Laplacian matrix $L$ and its grounded Laplacian matrix $\bar L$, of an undirected minimum $c^{\prime}$-algebraic connected graph family  $\{\mathcal{G}_N\}$, then, there exists a network size $\tilde N=1+\frac{2d_{{\max}}}{{c^{\prime}}^2d_{\min}}$ such that  $\frac{\lambda_2}{\lambda_N}>\frac{\bar \lambda_1}{\bar \lambda_{N-1}}$, for $N>\tilde N$.
\elemma
\begin{Proof}
By the eigenvalue interlacing theorem \cite{haemers1995interlacing}, $\lambda_N\geq\bar \lambda_{N-1}\geq\lambda_{N-1}\geq\dots\geq \lambda_2\geq c^{\prime}$, and since $\lambda_N\leq 2$, there exists $\tilde N=1+\frac{2d_{{\max}}}{{c^{\prime}}^2d_{\min}}$, such that for $N>\tilde N$ $$\frac{\lambda_2}{\lambda_N}\geq\frac{c'}{2}=\frac{d_{{\max}}}{c'(\tilde N-1)d_{\min}}>\frac{d_{{\max}}}{c'(N-1)d_{\min}}\geq\frac{\bar \lambda_1}{\bar \lambda_{N-1}},$$ 
where, in the last inequality we have used Lemma 4.2 and the fact that $N> \tilde{N} > \bar {N}$. The proof is completed.%thus the proof is completed. 
\end{Proof}
As seen from \eqref{Auinequal} in Assumption~\ref{asm2}, the eigenratio also characterizes the upper bound of allowable unstable margin for discrete-time system dynamics, i.e., ${\prod_j|\lambda_j^u(A)|}<\frac{1+\rho}{1-\rho}$. Grounding a network leads to a smaller eigenratio, then ``less unstable" system dynamics will be allowed. In the case that the unstable system dynamics exceed the consensusability upper bound after grounding, the consensusability of the whole network is lost.

%From \cite{yan2020analysis}, we know that grounding has an undesired impact on the network consensus in terms of scalability, consensus performance and consensusability. We recall the main results revealing the scaling fragility, consensus performance degradation and loss of consensusability relating to algebraic connectivity and eigenratio as follows. %(to be changed for non-regular graphs)

%

\section{Countermeasures}

In this section, we propose both passive and active countermeasures to preserve or recover the network properties after grounding. 

\subsection{Passive countermeasure} 
It is possible to design the controller beforehand to be resilient to grounding. At the stage of controller design, we select $K$ such that $(I_{N-1}\otimes A- \bar L\otimes BK)$ is Schur for every $\bar L$ resulting from grounding the $i{th}$ node, for $i=1,\dots,N$. Then, check that $K$ also stabilizes $( A-\lambda_i BK)$ for $i=2,\dots,N$. To guarantee the existence of such a grounding-resilient controller, Assumption~\ref{Auinequal} needs to hold with the eigenratio $\rho$ being replaced by the smallest grounded eigenratio among all those resulting from grounding the $i$th node, $i=1,\dots,N$.
%The necessary and sufficient condition for the existence of such a common control gain $K$ is given in the following lemma.%Such a $K$ is expected to exist by noting $\bar \lambda_1<\lambda_2\leq\cdots \leq \bar \lambda_{N-1}\leq \lambda_N$, the interlacing relationship of $\lambda_i$ and $\bar \lambda_i$, stated by the following.
% \blemma
%Given $0<\bar\lambda_1\leq\lambda_2\leq\dots\leq \lambda_N$, a necessary and sufficient condition for the existence of a common control gain $K\in \R^{1\times n}$ such that $\rho(A-\mu_mBK)<1$ for all $\mu_m\in\{\bar \lambda_{i-1},\lambda_i, i=2,\dots,N\}$ is the condition in Assumption~\ref{asm2} with \eqref{Auinequal} being replaced by
% \EQ
%\tilde \sigma>\frac{1-\tilde \rho}{1+\tilde \rho},~~\tilde\rho=\bar\lambda_1/\lambda_N.%=\frac{1}{\prod_j|\lambda_j^u(A)|}
%\EN
%\elemma%\leq\sigma
%\begin{Proof}
%
%\end{Proof}
%This technique may be only practical for networks of small size. 
\subsection{Active countermeasures} 
It is possible to take the following actions:

\begin{enumerate}
\item Suppose the grounded network is consensusable with respect to the agent dynamics. Then, redesign the controller after grounding such that $(I_{N-1}\otimes A-\bar L\otimes BK)$  is Schur. Furthermore, if the grounded algebraic connectivity is lower bounded by $c^{\prime}_g$ and $c^{\prime}_g>\frac{2(1-\tilde \sigma)}{1+\tilde \sigma}$, then, consensus can be achieved by \eqref{K} with $\sigma$ and $\epsilon$ selected in \eqref{sigma} and \eqref{epsilon} by replacing $c'$ with $c'_g$.

\item Suppose the grounded network is unconsensusable with respect to the agent dynamics and the grounded node can be detected. Then we can isolate the grounded node by cutting the connections between the grounded node with all its neighbors. If the rest of the network remains minimum $c^{\prime}$-algebraic connected, not only the consensusability can be recovered but also consensus can be achieved by the original non-grounded control design without reconfiguration. %The remaining network will still achieve consensus.

\item Suppose the grounded network is unconsensusable with respect to the agent dynamics. If the system dynamics are unstable, and consensusability is lost, then there does not exist a $K$ to stabilize $(I_{N-1}\otimes A-\bar L\otimes BK)$. Neither redesign nor predesign the controller for grounding will work in this case. We then propose a possible approach to regain the consensusability by deliberately grounding more nodes to increase the upper bounds for the allowable unstable dynamics. This may sound counter-intuitive, but it will be seen from the following Lemma~\ref{coro3} that by grounding more nodes, the grounded algebraic connectivity  increases (or does not decrease), and the spectral radius of the grounded Laplacian decreases (or does not increase), thus leading to a potential increase in eigenratio and larger allowable region for the dynamics to be unstable.

\end{enumerate}

\blemma \label{coro3}

Let $\bar L_{sym}^{(m)}$ be obtained by removing the first $m$ rows and the first $m$ columns of $L_{sym}$, $m\in\mathcal{Z}^+$, $0<m<N$, $\bar \lambda_1^{(m)}$ be the smallest eigenvalue of $\bar L_{sym}^{(m)}$, $\bar \lambda_{N-m}^{(m)}$ the largest eigenvalue of $\bar L_{sym}^{(m)}$. Let $q \in Z^+$ be such that $0<m<q<N$. Then, $\bar \lambda_1^{(m)}\leq\bar \lambda_1^{(q)}$, $\bar \lambda_{N-m}^{(m)}\geq\bar \lambda_{N-q}^{(q)}$.
\elemma
\begin{Proof}
 %Since $\bar L^{(m)}=(\bar L^{(m)})^T$ is a positive definite matrix, we have $\lambda_{\min}(\bar L^{(m)})\|y\|^2 < y^T\bar L^{(m)} y< \lambda_{\max}(\bar L^{(m)})\|y\|^2$. %We have%The smallest and largest eigenvalue of $\bar L^m$ can be depicted as the following optimization expression,
 By the Ritz-Rayleigh theorem, we have \EQ
 \lambda_{\min}(\bar L^{(m)})&=&\displaystyle \min_{y_1=\dots=y_m=0}\frac{y^TL_{sym}y}{y^Ty}\nonumber\\
 &\leq& \displaystyle \min_{y_1=\dots=y_q=0}\frac{y^TL_{sym}y}{y^Ty}=\lambda_{\min}(\bar L_{sym}^{(q)})\\
 \lambda_{\max}(\bar L^{(m)})&=&\displaystyle \max_{y_1=\dots=y_m=0}\frac{y^TL_{sym}y}{y^Ty}\nonumber\\&\geq& \displaystyle \max_{y_1=\dots=y_q=0}\frac{y^TL_{sym}y}{y^Ty}=\lambda_{\max}(\bar L^{(q)})
 \EN  where $y_j$ represents the $j$th element of $y$. Thus the proof is completed.
%With more nodes grounded, the constraints of the $\min y^TLy$ and $\max y^TLy$ become restricter, therefore $\min y^TLy$ will not decrease and $\max y^TLy$ will not increase.
 \end{Proof}
 Lemma~\ref{coro3} considered, without loss of generality, the first $m$ (or $q$) rows/columns but the results hold for any group of $m$ (or $q$). 
 
 It is also worth mentioning that the result in Lemma~\ref{coro3} can be obtained by noticing the eigenvalue interlacing theorem. We provide the self-contained proof here for the convenience of the readers. 
%\bremark
%%The proof of Corollary~\ref{coro3} can be completed noticing the eigenvalue interlacing theorem and is thus omitted.
%
%%Also, 
%Extensive simulation studies suggest that the strict inequalities generally hold, that is, in all cases examined, $\bar \lambda_1^{(m)} < \bar \lambda_1^{(q)}$, $\bar \lambda_{N-m}^{(m)}>\bar\lambda_{N-q}^{(q)}$. 
%\eremark

%All the above results on the grounded eigenvalue and grounded eigenratio can be summarized in the following general result for non-regular communication graphs with $m$ nodes grounded by giving the upper bound of the grounded algebraic connectivity $\bar \lambda_1^{(m)}$ corresponding to the grounded normalized Laplacian $\bar L^{(m)}$. Let $d_{\max}$ and $d_{\min}$ be the maximum and minimum degree, respectively, among all the $N$ nodes.
When the network has $m$ nodes grounded, the upper bound of the smallest eigenvalue of the resulting grounded Laplacian $\bar L^{(m)}$ can be obtained by straightforwardly extending Lemma~\ref{lemma_1_bound} as folows.
\bcorollary\label{mground}
  \label{bound}
  For a given $N$ and $m$, the smallest eigenvalue of the grounded Laplacian is bounded such that
  \begin{equation}\label{boundm}
    \bar{\lambda}_{1}^{(m)} \leq \frac{m ~{d}_{\max}}{(N-m) ~d_{\min} }
  \end{equation}
\ecorollary
\begin{Proof}
The proof is similar to the one for Lemma~\ref{lemma_1_bound} considering one node grounded. We note that when grounding $m$ nodes, the Laplacian $L$ can be partitioned as 
\EQ
L=\left[\begin{array}{cc}
L_0&\hat L\\
\tilde L& \bar L^{(m)}
\end{array}\right].
\EN
Consider $s=\left[\sqrt{d_{m+1}}, \dots,~\sqrt{d_N}\right]^T$ and $s^Ts=\sum_{i=m+1}^Nd_i$. Then, $s^T\bar L^{(m)}_{sym}s=s^T\sqrt{-\tilde L}\mathbf{1}_m$, $s^T\sqrt{-\tilde L}=[d_1 \dots d_m]$, and $$s^T\bar L^{(m)}_{sym}s=\sum_{i=1}^md_i\leq m\cdot \max{\{d_1,\dots,d_m\}}\leq md_{\max}.$$ Thus, \eqref{boundm} follows directly.
\end{Proof}
\bremark
In Corollary~\ref{mground}, as $m$ increases for a given network, the upper bound of the resulting grounded algebraic connectivity increases. This indicates that additional grounding can provide an improvement. However, it ultimately shows that as $N$ grows, more and more nodes need to be grounded for recovery. 
\eremark

\section{The Selection of Additional Nodes to Ground}
%We proposed an approach to regain the consensusability or recover the consensus performance in terms of convergence rate by deliberately grounding more nodes to increase the upper bounds for the allowable unstable dynamics or to increase the grounded algebraic connectivity. This may sound counter-intuitive, but it will be seen from the eigenvalue interlacing theorem that by grounding more nodes, the grounded algebraic connectivity increases (or does not decrease), and the spectral radius of the grounded Laplacian decreases (or does not increase), thus leading to a potentially increased eigenratio and larger allowable region for the dynamics to be unstable.

In this section, we investigate the questions of the proper selection of an additional node to be grounded and how many nodes we need to ground for recovery. 

\subsection{Which additional node should be grounded}
The selection of an additional node to ground is here aimed at achieving the maximal increase of the grounded algebraic connectivity. It is worth mentioning that when the consensusability of a network is lost after grounding, the additional node to ground should be chosen to attempt to recover the eigenratio. However, when the grounded network is still consensusable, then the additional node to ground can be selected to achieve the highest grounded algebraic connectivity to improve the scalability and consensus performance. 

In this section, we thus focus on how to choose an additional node from nodes $2$ to $N$ to be grounded to yield a maximal increase of the grounded algebraic connectivity for $c^{\prime}$-expander $d$-regular graphs. This problem can be formulated as follows:
\begin{equation}\label{maxmin}
\arg\!\!\!\max_{j=1,\dots,N-1}  \{ \lambda_{\min}(\bar L(j))\}
\end{equation}
where $\bar L(j)\in \R^{(N-1)\times(N-1)}$ denotes the double grounded Laplacian and is obtained by replacing the $j$th row and column of $\bar L$ by zeros, which also means by replacing the $(j+1)$th row and column of $L$  by zeros after grounding the first node. Since the graphs considered are $d$-regular, then $L$, $\mathcal{A}$ and their principal sub-matrices are all symmetric. Let the grounded adjacency matrix be $\bar{\mathcal{A}}$ obtained by removing the first row and the first column of $\mathcal{A}$, $\bar{\mathcal{A}}(j)$ obtained by replacing the $j$th row and column of $\bar{\mathcal{A}}$ by zeros. One has $\bar L=I_{N-1}-\bar {\mathcal{A}}$, $\bar L(j)=I_{N-1}-e_je_j^T-\bar{\mathcal{A}}(j)$, where $e_j\in\R^{N-1}$ denotes a column vector with all elements being $0$ except the $j$th element being $1$.
 The eigenvalues of $\bar L(j)$ are denoted by $0=\bar\lambda_0^{\prime}<\bar\lambda_1^{\prime}\leq\bar\lambda_2^{\prime}\leq \cdots \leq \bar\lambda_{N-2}^{\prime}$,
the eigenratio by $\bar\rho^{\prime}=\frac{\bar\lambda_1^{\prime}}{\bar\lambda_{N-2}^{\prime}}$. 
% $\varsigma(\boldsymbol{\cdot})$ denotes the spectrum of a square matrix $\boldsymbol{\cdot}$.

%To increase the practicality of the problem we now provide some\{ \lambda_{\min}(\bar L(j))\}

%We now provide some theoretical analysis on why the worst node which is with the minimal $\bar\lambda_1^{\prime}$ appearing to be a neighboring node.
% First
%We first denote the node that should be additionally grounded as the "recovering node", specifically, after grounding the recovering node the highest eigenratio can be achieved compared with additionally grounding other nodes.  As in \cite{lountzi2015expanders}, a node is said to be in the $i$-th layer if its shortest path to the grounded node $v_g$ has length $i$. A node is said to be in the highest layer if its shortest path to the grounded node $v_g$ has longest length.

The `max-min' problem described in \eqref{maxmin} appears to be hard as the solution relates to the structure of the double grounded Laplacian $\bar L(j)$. We then seek practical ways to select effectively the best node to ground without checking all the $\lambda_{\min}(\bar L(j))$ for $j=1,\dots,N-1$. To this end, we first define that a node is in the $i$-th layer if its shortest path to the grounded node has length $i$. A node is said to be in the highest layer if its shortest path to the grounded node has the longest length, with $\ell$ being the longest length. We start by reordering nodes by layers resulting in the following reordered adjacency matrix,
\EQ
{\mathcal{A}}_{ord}=\left.\begin{array}{cccccc} 0\mbox{th}\\1\mbox{st}\\2\mbox{nd}\\\vdots\\(\ell-1)\mbox{th}\\\ell \mbox{th} \end{array}\right.\left[
\begin{array}{c|c|c|c|c|c}
0 &\frac{\mathbf{1}_d^T}{d} &0 &0&0 &0 \\ \hline
\frac{\mathbf{1}_d}{d}&\mathcal{A}_1&\mathcal{S}_{12}&0&0&0 \\ \hline
0&\mathcal{S}_{12}^T&\mathcal{A}_2 &\mathcal{S}_{23}&0&0\\ \hline
0&0&\mathcal{S}_{23}^T&\ddots&\vdots&0\\ \hline
0&0&0&\cdots&\mathcal{A}_{\ell-1}&\mathcal{S}_{\ell-1, \ell}\\\hline
0&0&0&0&\mathcal{S}_{\ell-1, \ell}^T&\mathcal{A}_{\ell}\\
\end{array}\right]
\EN
where $1_n$ denotes  the column vector of $n$ ones, $\mathcal{A}_j$, $j=1,\dots,\ell$, denotes the interconnection of nodes in the $j$th layer, $\mathcal{S}_{j,j+1}$ represents the interconnection between nodes in the $j$th layer and those in the $(j+1)$th layer.
In this permutation, ${\mathcal{A}}_{{ord}}$ is block tri-diagonal. After grounding of node $1$ occurs, the grounded reordered adjacency matrix $\bar{\mathcal{A}}_{{ord}}$ is obtained by removing the first row and the first column of $\mathcal{A}_{ord}$. One has
\EQ\label{Aord}
\mathcal{A}_{ord}\mathbf{1}_{N}=\mathbf{1}_{N},~\bar{\mathcal{A}}_{ord}\mathbf{1}_{N-1}=\mathbf{1}_{N-1}-\left[\begin{array}{cccc}\frac{\mathbf{1}_{d}}{d}\\
0\\
\vdots\\
0\\
\end{array}\right].
\EN

Consider the vector $v_1=\mathbf{1}_{N-1}$ and $\bar{\mathcal{A}}_{ord}(j)$ denotes the double grounded adjacency matrix obtained by replacing the $j$th row and the $j$th column of $\bar{\mathcal{A}}_{ord}$ by zeros. Then, the maximum eigenvalue of the reordered adjacency matrix can be bounded as 
\EQ
&&\lambda_{\max}(\bar{\mathcal{A}}_{ord}(j)):=\max_{v\neq 0}\frac{v^T\bar{\mathcal{A}}_{ord}(j)v}{v^Tv}\nonumber\\&&\geq \frac{v_1^T\bar{\mathcal{A}}_{ord}(j) v_1}{v_1^Tv_1}\\&&=\frac{v_1^T(\bar{\mathcal{A}}_{ord}-e_je_j^T\bar{\mathcal{A}}_{ord}-\bar{\mathcal{A}}_{ord}e_je_j^T)v_1}{N-1}\label{right_ineq}
\EN
Note that
 \EQ\bar\lambda^{\prime}_1&=&\lambda_{\min}(\bar L(j))=1-\lambda_{\max}(\bar{\mathcal{A}}(j)).\EN
Minimising the lower bound \eqref{right_ineq} will allow more `room' for minimising $\lambda_{\max}(\bar{\mathcal{A}}_{ord}(j))$ thus maximising $\lambda_{\min}(\bar{L}(j))$ as $\lambda_{\max}(\bar{\mathcal{A}}(j))=\lambda_{\max}(\bar{\mathcal{A}}_{ord}(j))$. Selecting $j=j^*$ such that
 $$j^*=\arg\max_{j=1,\dots,N-1}~~v_1^T(e_je_j^T\bar{\mathcal{A}}_{ord}+\bar{\mathcal{A}}_{ord}e_je_j^T)v_1,$$
 then the lower bound can be minimised. Noting that $\psi_j=v_1^T(e_je_j^T\bar{\mathcal{A}}_{ord}+\bar{\mathcal{A}}_{ord}e_je_j^T)v_1$ is the addition of the $j$th row sum and $j$th column sum, $\psi_{j\in\mathcal{N}_1}$ is less than $\psi_{j\in \mathcal{V}/{\mathcal{N}}_1}$, since, from \eqref{Aord}, $\psi_{j\in\mathcal{N}_1}-\psi_{j\in\mathcal{V}/{\mathcal{N}}_1}=\frac{-2}{d}$. Then, we conclude that the neighboring nodes are worse candidates to ground than non-neighboring nodes when considering the available `room’  to recover the grounded eigenvalue. This suggests that the selection of the additional node to be grounded relates to the distance from the first grounded node, namely, node $1$. In fact, Monte Carlos simulations suggest that the best additional node to ground usually appears in the highest two layers $\ell$ and $\ell-1$. Motivated by these findings, we propose the following algorithm to assist us in selecting an additional node to ground aiming at recovering the algebraic connectivity.

%for $j\notin \mathcal{N}_i$. the addition for node $j$ which is the neighboring node is $2$ less than that for other nodes. This explains 

%\bru
   \begin{algorithm}[!ht]
          \caption{} \label{conjecture}
For a $c^{\prime}$-expander $d$-regular graph family $\{\mathcal{G}_N\}$, the additional node towards achieving maximal increase of the grounded algebraic connectivity should be selected from the set of nodes in the highest two layers with respect to the grounded node.% $v_g$.%have the longest shortest path from the first grounded node.% if such a node will not disconnect the graphs.
%\eru
\end{algorithm}

\begin{figure}[ht!]
\begin{center}
\includegraphics[width=8cm]{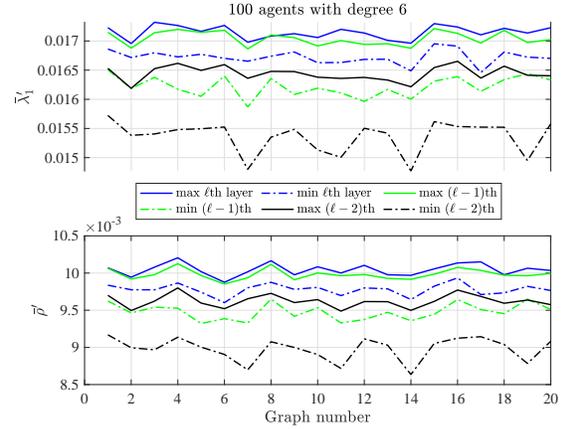}   
\caption{The max/min values of the double grounded algebraic connectivity $\bar\lambda_1^{\prime}$ and the double grounded eigenratio $\bar \rho^\prime$ in different layers, $\ell=4$.} 
\label{fig:range}
\end{center}
\end{figure}

\begin{figure}[ht!]
\begin{center}
\includegraphics[width=8cm]{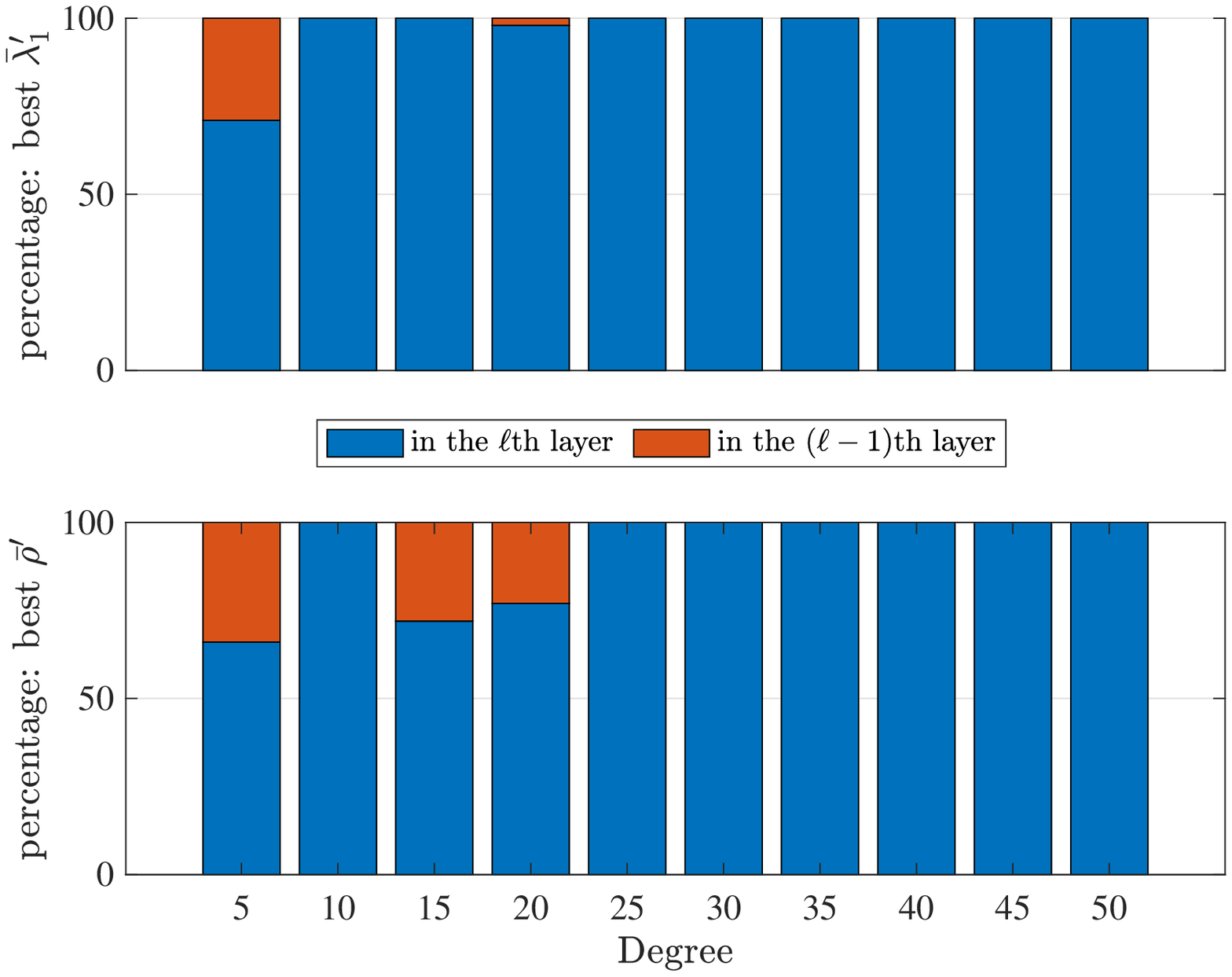}   
\caption{The percentage that the double grounded algebraic connectivity $\bar\lambda_1^{\prime}$ and the double grounded eigenratio $\bar \rho^\prime$ are in the highest or 2nd highest layers, $N=100$.} 
\label{fig:percentage2}
\end{center}
\end{figure}

\begin{figure}[ht!]
\begin{center}
\includegraphics[width=8cm]{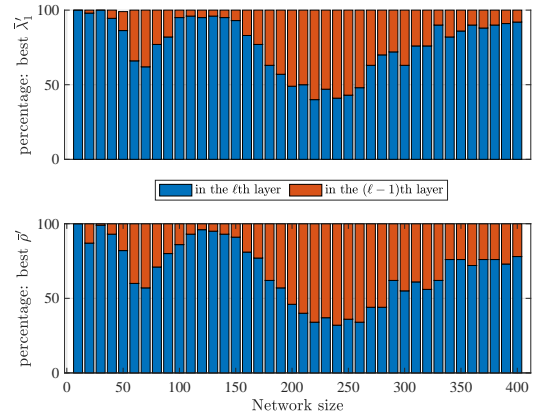}   
\caption{The percentage that the double grounded algebraic connectivity $\bar\lambda_1^{\prime}$ and the double grounded eigenratio $\bar \rho^\prime$ are in the highest or 2nd highest layers, $d=6$.}% (no pattern here, just for your reference and will be removed.)} 
\label{fig:percentage1}
\end{center}
\end{figure}

We illustrate this algorithm by the following example as seen in Fig.~\ref{fig:range}. This figure shows the distribution of the double grounded algebraic connectivity $\bar\lambda_1^{\prime}$ and double grounded eigenratio $\bar \rho^\prime$. To produce this figure, we use the algorithm in \cite{kim2003generating} to generate $20$ graphs of $100$ agents with degree $6$. For each graph and each layer, we plot the maximum and minimum values of the double grounded algebraic connectivity (top plot) and the maximum and minimum values of the double grounded eigenratio (bottom plot).  We can see that the range of $\bar\lambda_1^{\prime}$ appears to be in `belt'-shape in terms of which layer the node is located. The figure shows that the worst node, that is the node with the minimal $\bar\lambda_1^{\prime}$ is always a neighboring node of the grounded node. The best node
 appears most likely  in the highest layer $\ell$, and sometimes in the layer $\ell-1$. If we change the network by allowing different degrees, the percentage of the best node in terms of  $\bar\lambda_1^{\prime}$ and $\bar \rho^\prime$ appearing in the $\ell$th layer or in the ($\ell-1$)th layer can be seen in Fig.\ref{fig:percentage2}. This was obtained by randomly generating $100$ graphs using the algorithm in  \cite{kim2003generating}  for each case.  We fix the network size $N=100$ and have the degree changing as $d=5, 15,\dots,50$. When the degree increases up to half of the network size, the best node is almost always in the highest layer. Fig.~\ref{fig:percentage1} shows the percentage of the best node appearing in each layer when we fix the degree $d=6$, and change $N$ from $10, 20,\dots, 400$. %we illustrate that when the best node does not appear in the highest layer but in the 2nd highest layer, it is because usually the highest layer is $1$ length higher than the normal case and there are very few nodes in the highest layer (which was not shown). 
%~\ref{fig:percentage1} and Fig.~\ref{fig:percentage1} shows that when we fix the degree $d=6$, and change the $N$ from $10, 20,\dots, 100$, .Assuming $d\geq 5$ is to exclude the case of non-expanders after grounding two nodes. 

Next,  we recall Perron-Frobenius theorem  for non-negative matrices (with all the entries non-negative).
 \btheorem(\cite{horn2012matrix})
 Let $T\in\R^{n\times n}$ be an irreducible non-negative matrix, then,
 \begin{itemize}
 \item $T$ has a positive (real) eigenvalue $\lambda_{\max}$ such that all other eigenvalues of T satisfy $|\lambda|<\lambda_{max}$;
 \item The left and right eigenvector corresponding to the eigenvalue $\lambda_{\max}$ have all positive components.
 \end{itemize}
 \etheorem
 Consider $\bar{\mathcal{A}}$, which is non-negative, and assume the remaining network (excluding node $1$) is strongly connected. Then $\lambda_{\max}(\bar{\mathcal{A}})$ is real and unique, and its corresponding normalized eigenvector $v_{\max}$ has positive elements with $v_{\max}^T\textbf{1}_{N-1}=1$.  Similar to the analysis leading to Algorithm~\ref{conjecture}, we are interested in choosing $j$ to minimise $\lambda_{\max}(\bar{\mathcal{A}}(j))$. One has
 \EQ 
 &&\lambda_{\max}(\bar{\mathcal{A}}(j)) \geq\frac{ {v}_{\max}^T\bar{\mathcal{A}}(j){v}_{max}}{{v}_{max}^T{v}_{\max}}\nonumber\\&=&\frac{ {v}_{\max}^T(\bar{\mathcal{A}}-e_je_j^T\bar{\mathcal{A}}-\bar{\mathcal{A}}e_je_j^T){v}_{max}}{{v}_{max}^T{v}_{\max}},
% &=& \lambda_{\max}(\bar{\mathcal{A}})-\frac{2(v_{\max}^Te_je_j^T\bar{\mathcal{A}})v_{\max, j}}{v_{\max}^Tv_{\max}}
 \EN 
% where $\bar{v}_{\max}$ is obtained by removing the $j$th element of $v_{\max}$, 
then,
  \EQ\label{inequality}
 \lambda_{\max}(\bar{\mathcal{A}}(j))\geq  \lambda_{\max}(\bar{\mathcal{A}})-\frac{2(v_{\max}^T\bar{\mathcal{A}}e_j)v_{\max, j}}{v_{\max}^Tv_{\max}},
 \EN
 where  $v_{\max,j}$ represents the $j$th element of $v_{\max}$. Let $\tilde\psi_j=[(v_{\max}^T\bar{\mathcal{A}}e_j)v_{\max, j}]$. Selecting $j=j^*$ such that $j^*=\arg\max_{j=1,\dots,N-1}\tilde \psi_j$, the lower bound in \eqref{inequality} will be minimized. As discussed before, this gives more `room' for minimising $\lambda_{\max}(\bar{\mathcal{A}}(j))$ thus maximising $\lambda_{\min}(\bar{L}(j))$. Note that $\bar{\mathcal{A}}e_j$ represents the $j$th column of $\bar{\mathcal{A}}$ which contains mostly $0$ entries if the network is sparse. In this case, the major contribution comes from choosing the maximal $v_{\max,j}$. We can see that $\bar{\mathcal{A}}^k \mathbf{1}_{N-1}$ approaches $v_{\max}$, as $k\to \infty$. Thus we propose to choose the maximum element of $\bar{\mathcal{A}}^{\ell} \mathbf{1}_{N-1}$ as an approximation of $v_{\max, j}$. Based on above analysis, we propose the following algorithm to select the best node to ground.
%{\blue It can be seen that $\lambda_{\max}(\bar{\mathcal{A}}(j))$ is lower bounded by the right-hand side of \eqref{inequality}.  To allow more 'room' for minimising $\lambda_{\max}(\bar{\mathcal{A}}(j))$ thus maximising $\lambda_{\min}(\bar{L}(j))$, we are interested to minimise this lower bound. 
   \begin{algorithm}[!ht]
          \caption{} \label{alg}
%\bru\label{alg} 
%Calculate the $\ell$th power of $\bar{\mathcal{A}}$ and compute its row/column sums, which is equivalent to compute $\bar{\mathcal{A}}^{\ell}{\bf{1}}_{N-1}$. Then, 
For a $c^{\prime}$-expander $d$-regular graph family $\{\mathcal{G}_N\}$, the additional node towards achieving the maximal increase of the grounded algebraic connectivity should be selected from the set of nodes that yield the largest element of %sums (largest density)
$\bar{\mathcal{A}}^{\ell}{\bf{1}}_{N-1}$.% are almost the best nodes to ground. 
\end{algorithm}
%\eru
%Algorithm 1:
%\begin{enumerate}
%\item  Let the grounded adjacency matrix be $\bar{\mathcal{A}}$, which is obtained by removing the first row and first column of the adjacency matrix. 
%\item Calculate the $\ell$th power of $\bar{\mathcal{A}}$ and compute its row/column sums, which is equivalent to compute $\bar{\mathcal{A}}^{\ell}{\bf{1}}_{N-1}$.% where $\ell$ denotes the longest shortest path from the grounded node to the other nodes, $[\mathcal{A}_g^{\ell}]_j$ denotes the $j$th row. 
%
%\item Then, the nodes corresponding to the largest element of %sums (largest density)
%$\bar{\mathcal{A}}^{\ell}{\bf{1}}_{N-1}$ are almost the best nodes to ground. 
%
%\end{enumerate} 
   
\begin{figure}[ht!]
\begin{center}
\includegraphics[width=8cm]{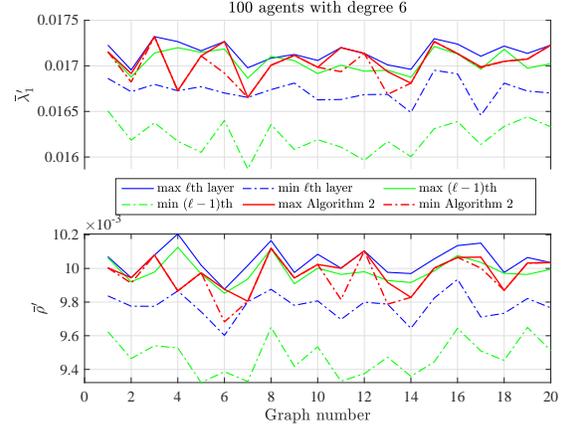}   
\caption{The estimation given by Algorithm~\ref{alg} is close to the best node, $\ell=4$.} 
\label{fig:compare100_6}
\end{center}
\end{figure}

It is noted that Algorithm~\ref{alg} essentially picks $j$ to maximise an approximation of $v_{\max, j}$.  Fig.~\ref{fig:compare100_6} validates the usefulness of Algorithm~\ref{alg} by showing that the smallest eigenvalue $\bar\lambda_1^{\prime}$ of the resulting double grounded matrix $\bar L(j)$ corresponding to the selected nodes is very close to that corresponding to the best nodes. Note that Fig.~\ref{fig:compare100_6} is generated in the same way as Fig.~\ref{fig:range}, with the addition of the selection achieved by Algorithm~\ref{alg}.

\begin{figure}[ht!]
\begin{center}
\includegraphics[width=8cm]{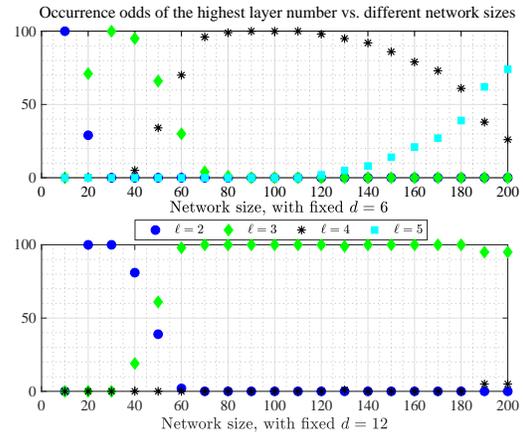}   
\caption{The percentage of $\ell$ with different network size } 
\label{fig:ell}
\end{center}
\end{figure}

Finally, we observe that, $\ell$ is usually not very large ($<10$) as seen in Fig.~\ref{fig:ell}, which shows the number of the highest layer $\ell$ as a function of network size for two different degrees. For each network size, we generate $100$ random graphs and record the distribution of $\ell$. It is noted that $\ell$ grows very slowly with $N$, for example, $\ell\leq 5$ for $N\leq 200$. We also recorded that when $N=1000, d=6$, $\ell=6$ for all the $100$ randomly generated graphs.

\subsection{How many nodes should be grounded}
%We now focus on the question of how many nodes we need to ground to recover the eigenvalue to be larger or equal to the original non-grounded algebraic connectivity $\lambda_2$. 

In \cite{yan2020analysis}, it was pointed out that it is always possible to recover consensusability by grounding a sufficiently large number of additional nodes for regular expanders by using a worst case estimation analysis. 
%Consider expanders with $d\geq 3$, when altogether we ground $N-2$ nodes in an extreme case, $\frac{\bar \lambda^{(m)}}{\bar \lambda_N^{(m)}}\geq\frac{1}{2}>\frac{\lambda_2}{\lambda_N}$. Then, $\prod_j|\lambda_j^u(A)|<\frac{1+\bar\lambda_1^{(m)}/\bar\lambda_{N-m}^{(m)}}{1-\bar\lambda_1^{(m)}/\bar\lambda_{N-m}^{(m)}}$ which implies that the consensusability is recovered. 

%Note that the above considers a worst case estimation and in our simulations 
Simulations indicate that usually after randomly grounding approximately half of the nodes the ``pre-grounding'' consensusability can be recovered. Here the ``pre-grounding'' consensusability is recovered in the sense that the grounded eigenratio becomes very close to the original eigenratio $\rho$. However, grounding only a \emph{few} additional nodes $(<5)$ usually recovers consensusability. In particular, \eqref{Auinequal} in Assumption~\ref{asm2} restricts the system dynamics to be $\prod_j|\lambda_j^u(A)|<\frac{1+\rho}{1-\rho}$. If there is an $m$ such that $\frac{1+\bar\rho^{(m)}}{1-\bar \rho^{(m)}}>\prod_j|\lambda_j^u(A)|$, the network will be consensusable again. 

Essentially how many nodes to ground for regaining consensusability will be based on how unstable the system dynamics are compared to the upper bound in terms of the eigenratio. The closer $\prod_j|\lambda_j^u(A)|$ is to $\frac{1+\rho}{1-\rho}$, the more nodes will be needed to ground. It is possible that proper selection of the nodes to be grounded can reduce the necessary number. %Detailed analysis of these matters will be discussed in the following.

\begin{figure}[h]
\begin{center}
\includegraphics[width=8cm]{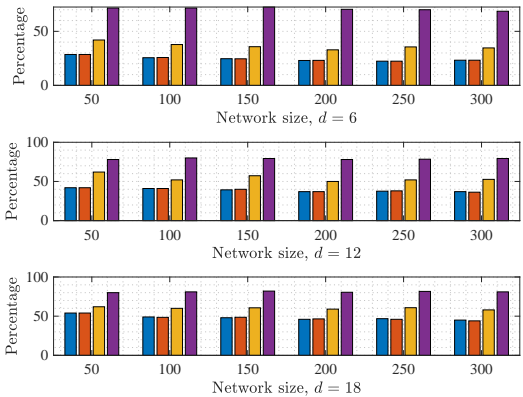}   
\caption{Percentage of nodes that need to be grounded to recover algebraic connectivity: From left to right, 1) blue: best direction, 2) red: Algorithm~\ref{alg}, 3) yellow: random direction, 4) purple: worst direction.} 
\label{fig:many}
\end{center}
\end{figure}

In relation to recovering the grounded algebraic connectivity so that it is larger or equal to the non-grounded algebraic connectivity, we next explore the use of Algorithm~\ref{alg} to guide the selection of nodes to ground. In Fig.~\ref{fig:many}, we plot the percentage of nodes that need to be grounded to recover the algebraic connectivity when the nodes are selected in four different ways. The first one is along the `best direction', which means to select additional nodes with maximal grounded eigenvalue every time by comparing the eigenvalues of all the possible grounded Laplacians. The second one is to select additional nodes using Algorithm~\ref{alg}, i.e., to find the node with the largest element in $\bar{\mathcal{A}}^{\ell}{\bf{1}}_{N-1}$. The third one is by randomly grounding extra nodes; here for example, we select the nodes in the order of natural numbers (to ground $m$ extra nodes is to ground nodes $2$ to $m+1$). The last one is along the `worst direction’, which means to select additional nodes with minimal grounded algebraic connectivity every time. The bars in the figure indicate the percentage of nodes that need to be grounded to recover the algebraic connectivity. We compute the average percentage by randomly generating ten regular graphs. As expected, along the best direction we need to ground the smallest percentage of nodes while the largest percentage is needed along the worst direction and somewhere in between for random direction. It can be seen that the percentage by using the proposed Algorithm~\ref{alg} is almost identical to that for the best direction, which illustrates the effectiveness of the proposed algorithm. 

\begin{figure}[h]
\begin{center}
\includegraphics[width=8cm]{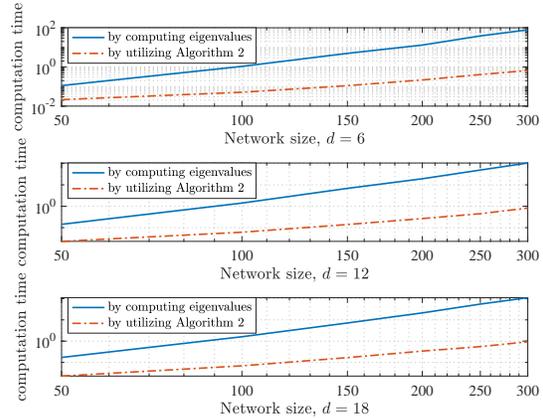}   
\caption{Computation time (unit: second) by computing the eigenvalues or by utilizing Algorithm~\ref{alg}. } 
\label{fig:computation_time}
\end{center}
\end{figure}

Moreover, we emphasize that Algorithm~\ref{alg} uses much less computational load than computing and comparing the eigenvalues directly. In fact, the computational load of Algorithm~\ref{alg} is  $\mathcal{O}(N^{2.373} \log({\ell}))$ with respect to the Coppersmith-Wingard algorithm \cite{coppersmith1987}. %(Coppersmith-Wingard algorithm used here, maybe it has faster complexity now, reference to be added.)
In Fig.~\ref{fig:computation_time}, we plot the computation time both for directly computing the eigenvalues and for utilizing Algorithm~\ref{alg}. The simulation platform is MATLAB on a laptop computer of 2.2 GHz Intel Core i7 and 16 GB memory. We can see that the computation time grows exponentially by computing eigenvalues when the network size and nodal degree grow and apparently
the proposed Algorithm~\ref{alg} consumes much less time than directly computing eigenvalues.

\begin{figure}[h]
\begin{center}
\includegraphics[width=8cm]{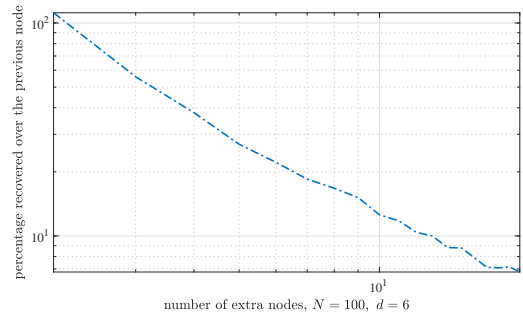}   
\caption{Percentage recovered compared to the recovery achieved by the previous node} 
\label{fig:percentage_recover}
\end{center}
\end{figure}

Recognizing that grounding many additional nodes may be not practical, it is important to emphasize that grounding the first additional node results in the largest incremental step towards recovery of the original algebraic connectivity. This is seen in Fig.~\ref{fig:percentage_recover}, where the vertical axis indicates the incremental increase ratio, for example, after grounding a second node, the incremental increase ratio with respect to the grounded algebraic connectivity $\bar \lambda_1$ is $\frac{\bar \lambda_1^{\prime}-\bar \lambda_1}{\bar\lambda_1}$.  The more nodes we ground, the less pronounced the recovery influence is.
%which has more than doubled increment compared to the grounded eigenvalue

\section{Illustrations of Network Performances} 
\subsection{Lack of scalability over grounding}\label{Sec6.1}

Consider a leaderless vehicle platoon where the dynamics of each vehicle is modeled as a discrete-time double integrator, 
\EQ\label{vehicle}
x_{i1}(k+1)&=&x_{i1}(k)+x_{i2}(k)\nonumber\\
x_{i2}(k+1)&=&x_{i2}(k)+u_i(k),~i=1,\dots,N
\EN
In~\eqref{vehicle}, $x_{i1}$ denotes the position from a desired setpoint, $x_{i2}$ the velocity, $u_i$ the control input of the $i^{th}$ vehicle. The system takes the form of \eqref{sys1} with $A=\left[\begin{array}{cc}1&1\\0&1 \end{array} \right]$, $B=\left[\begin{array}{cc}0\\1 \end{array} \right]$. The consensus objective is to have the string of vehicles travel while maintaining a certain formation, e.g., a constant target inter vehicle spacing in this example. The reference trajectory can be described as $x^*_1(k)=x^*_2\cdot k+\delta_i(k)$ with a constant speed $x^*_2=1$ and constant spacing $\delta_i(k)=5$.

\begin{figure}[ht!]
\begin{center}
\includegraphics[width=8.4cm]{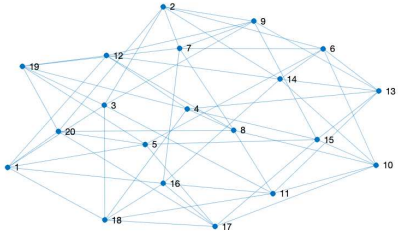}   
\caption{A random $6$-regular graph with $N=20$.} 
\label{fig:graph_regular}
\end{center}
\end{figure}

The communication graph family is minimum $0.3$-algebraic connected and is generated using the algorithm in \cite{kim2003generating} with two cases, $N=20$ and $N=100$, both with degree $d=6$. See for example in Fig.~\ref{fig:graph_regular} with $20$ nodes.

\begin{figure}[ht!]
\begin{center}
\includegraphics[width=\hsize]{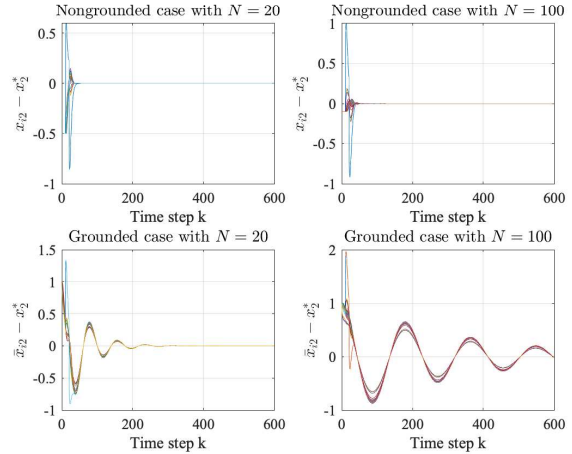}    %x20_100_0412
\caption{Profiles of velocity state deviations in nongrounded (leaderless) and grounded case (leader-following) with small and large network, perturbed by a sudden acceleration of one vehicle.} 
\label{fig:x20_100}
\end{center}
\end{figure}

The control gain is chosen as $K= [0.1424~0.9065]$ with $R=1$, $c^{\prime}=0.3$, $\sigma=0.8$, and $\epsilon=0.85$. The results of the simulation are shown in Fig.~\ref{fig:x20_100} comparing the nongrounded case (leaderless) and the grounded case (leader-following) when relating the state to an independent reference system with dynamics $x ^*(k+1)=(A-BK)x^*(k)+Bc_1$, $c_1=1$, for two network sizes $N=20$ and $N=100$. During the time steps $10-20$, one of the vehicles accelerates at a doubled speed. This disturbance is attenuated by the network within a short period of time for the nongrounded network. In contrast, for the grounded networks, the disturbance is attenuated with long settling time for the smaller network and even longer for the larger one. This illustrates the scalability limitation over the grounded networks.

\subsection{Loss of consensusability over grounding}\label{Sec6.2}
\begin{figure}[ht!]
\begin{center}
\includegraphics[width=8.4cm]{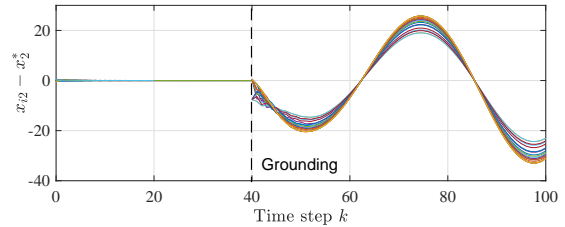}   
\caption{Profiles of state deviations: loss of consensusability after grounding. } 
\label{fig:grounding}
\end{center}
\end{figure}

Consider the consensus network~\eqref{sys1} with unstable dynamics $A=\left[\begin{array}{cc}1.07&1\\0&1 \end{array} \right]$, input matrix $B=\left[\begin{array}{cc}0\\1 \end{array} \right]$. The communication graph is as in the previous example with $N=20$. By properly designing the controller \eqref{con} with $K=[0.1777,~0.9649]$, where $\sigma=0.8$, and $\epsilon=0.85$, the consensus is achieved before $k=40$. At $k=40$, one of the agents is grounded, as illustrated in Fig.~\ref{fig:grounding}. Then, the network becomes unconsensusable since the consensusability condition~\eqref{Auinequal} is not satisfied after grounding. More specifically, $\bar \delta_A=\frac{1-\rho}{1+\rho}=\frac{1}{1.0596}>\frac{1}{\prod_{j}\left|\lambda_{j}^{u}(A)\right|}=\frac{1}{1.07}> \delta_A=\frac{1-\rho}{1+\rho}=\frac{1}{1.6935}$.%, which means that the grounded case allows a less unstable $A$. $P=\left[\begin{array}{cc}1.1142 & 3.2460\\ 3.2460&15.6106\end{array} \right]$

%%%%%%%%%%%%%%%%%%%%%%%%%%%%%%%%%%%%%%%%%
\subsection{Countermeasure through grounding more nodes}
\begin{figure}[ht!]
\begin{center}
\includegraphics[width=9cm]{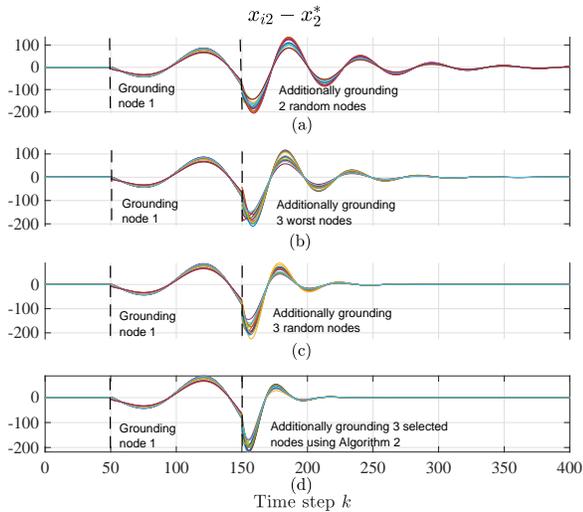}   
\caption{ Loss (at k=50) and regaining (after k=150) of consensusability after grounding: (a) randomly grounding $2$ additional nodes $(2, 3)$; (b) grounding $3$ worst nodes $(20, 19, 3)$; (c) randomly grounding $3$ additional nodes $(2, 3, 4)$; (d) grounding $3$ selected nodes $(13, 7, 17)$ using Algorithm~\ref{alg}.} 
\label{fig:grounding_more}
\end{center}
\end{figure}

Consider the same unstable network and communication graph as in Section~\ref{Sec6.2}. At $k=50$, node 1 is grounded, then as in Section~\ref{Sec6.2}, the consensusability is lost and the system states start to diverge. At $k=150$, we deliberately ground two more random nodes, $(2, 3)$. Then, the whole network will gradually achieve consensus again since the consensusability condition~\eqref{Auinequal} can be satisfied again. Furthermore, we compare this case with grounding $3$ additional nodes in three different directions, the worst direction, random grounding, and grounding selected nodes using Algorithm~\ref{alg}. We can see that the recovery convergence rate varies among four cases described above. The more nodes we ground, the faster the consensus performance is recovered. In addition, the proposed Algorithm~\ref{alg} helps to achieve the best recovery comparing to the random grounding as well as grounding along the worst direction. These findings are illustrated in Fig.~\ref{fig:grounding_more}. % $\prod_{j}\left|\lambda_{j}^{u}(A)\right|=1.07<\frac{1+\bar\rho^{(2)}}{1-\bar\rho^{(2)}} =1.1266$. 

%In an additional test, at $k=150$, instead of randomly grounding a second node (node $2$), %since $\bar \lambda_1^{(3)}=0.7026>\bar \lambda_1^{(2)}=0.5654$.we use Rule~\ref{alg} which gives us node $13$ to be grounded. It is observed that the consensus is achieved faster than the previous case. 

%%%%%%%%%%%%%%%%%%%%%%%%%%%%%
\section{Conclusion}
In this paper, we have analyzed the scaling fragility of expanders over grounding in a discrete-time context. As in a continuous-time setting, grounded expanders do not scale well. We give a proof that the eigenratio of the grounded network will approach zero, while the one of the nongrounded expander family is bounded away from zero. This shows that the consensus performance is deteriorated and in extreme cases can even lose consensusability. We give a condition under which the grounded network is able to achieve consensus.  In addition, we have discussed possible countermeasures for avoiding the loss of or regaining consensusability. The three methods discussed are to design the initial controller such that the grounded network remains stable, redesign the controller after grounding occurs, or deliberately ground additional nodes. We have also studied the problem of which additional nodes should be grounded so that the resulting network has the largest grounded algebraic connectivity and have proposed two algorithms to assist the selection. How to detect the grounded node will be investigated in future work.

%and then either adjust the controller
 \bibliographystyle{plain}%IEEEtran
\bibliography{Yaminsbib}

%\newpage
%
%\begin{center}
% \begin{tabular}{||c| c | c||} 
% \hline
%grounded nodes & grounded eigenvalue & grounded eigenratio \\ [0.5ex] 
% \hline\hline
% first node 1 &  0.2229 &   0.0446\\ 
% \hline\hline
% further node 2 & 0.4194 & 0.0839\\
% \hline
%  further node 3 & 0.3412 & 0.0691\\
% \hline
%  further node 4 &  0.4384 & 0.0877\\
%  \hline
%  further node 5 & 0.5188 &0.1038 \\
%  \hline
%  further node 6 &   0.3412 & 0.0691\\
%  \hline
% further node  7 &  0.4194&  0.0839\\
%  \hline
%  further node 8 & 0.2755& 0.0559\\
%  \hline
%  further node 9 & 0.3588&  0.0718 \\
%  \hline
%  further node 10 &  0.2755& 0.0559\\[1ex] 
% \hline
%\end{tabular}
%\end{center}

\end{document}